\documentclass[twocolumn,twoside,slac_two]{revtex4}
\usepackage{graphicx}
\usepackage{fancyhdr}
\newcommand{\Ptg}{p_{T}^{\gamma}}
\newcommand{\ptg}{p_T^{\gamma}}
\newcommand{\lt}{\!<\!}
\newcommand{\gt}{\!>\!}
\newcommand{\gpj}{$p\bar{p} \rightarrow \gamma + \mathrm{jet} + X$~}

\newcommand{\newscale}{p_T^\gamma f(y^\star)}
\newcommand{\Fystar}{\{[1+\exp(-2|y^\star|)]/2\}^{1/2}}
\newcommand{\ystar}{0.5(y^\gamma-y^{\text {jet}})}

\newcommand{\rJLIP}{P_{\rm HF-jet}}
\newcommand{\triplexs}{${{\rm d}^3\sigma/({\rm d}\Ptg {\rm d}y^\gamma {\rm d}y^{\rm jet})}$~}
\newcommand{\GeV}{{\rm GeV}~}
\newcommand{\gb}{$\gamma+{\rm b}+X$~}
\newcommand{\gc}{$\gamma+{\rm c}+X$~}

\newcommand{\DO}{D\O{}~}
\newcommand{\gpTHRj}{$\gamma+{\rm 3~jets}$~}

\newcommand{\ptsj}{p_T^{\rm jet2}}

\begin{document}

\title{Photon+jets measurements at D0}

%

\author{Dmitry V. Bandurin}
\affiliation{Kansas State University, Kansas, Manhattan, KS 66506, USA}

\begin{abstract}
In this paper, we present a few measurements done at the D0 experiment at the Fermilab
Tevatron Collider.
These measurements include
the triple differential cross sections ($d^3\sigma/dp_T^{\gamma}dy^\gamma dy^{jet}$) of the photon and associated jet
production, 
the photon and heavy flavour ($b$ and $c$) jet,  and finally, study of the event with double parton 
scattering using $\gamma+$3 jet events.
Each section below presents a brief description of those measurements and results.
\end{abstract}

\maketitle

\thispagestyle{fancy}


\section{Measurement of the differential cross section for the production of an isolated photon with
associated jet in $\bf p\bar{p}$ collisions at $\bf \sqrt{s}=$1.96 TeV}

\label{sec:gamjet}

The production of a photon with associated jets in the final state is a 
powerful probe of the dynamics of hard QCD interactions \cite{JFOwens}. 
Different angular configurations between the photon and the jets can be used to extend 
inclusive photon production measurements 
and simultaneously test the underlying dynamics of QCD hard-scattering 
subprocesses in different regions of parton momentum fraction $x$ and 
large hard-scattering scales $Q^2$.

Here we present an analysis of photon plus jets production in $p\bar{p}$ 
collisions at a center-of-mass energy $\sqrt{s}=$1.96 TeV in 
which the most-energetic (leading in $p_T$) photon is produced centrally with 
a rapidity $|y^{\gamma}|<1.0$ \cite{gamjet_PLB}. 
The cross section as a function of photon transverse 
momentum $\Ptg$ is measured differentially for four separate angular 
configurations of the highest $p_T$ (leading) jet and the 
leading photon rapidities. The leading jet is required to be in either the 
central ($|y^{\mathrm{jet}}|\lt 0.8$) or forward ($1.5\lt|y^{\mathrm{jet}}|\lt 2.5$) rapidity intervals, 
with $p_T^{\mathrm{jet}} > 15$ GeV. The four 
angular configurations studied are: central jets with $y^{\gamma}\! \cdot\! y^{\mathrm{jet}}>0$ 
and with $y^{\gamma} \cdot y^{\mathrm{jet}}<0$, and forward jets with 
$y^{\gamma} \cdot y^{\mathrm{jet}}>0$ and with $y^{\gamma} \cdot y^{\mathrm{jet}}<0$.
The total $x$ and $Q^{2}$ region covered by the measurement is $0.007\lesssim x \lesssim 0.8$ and 
$900 \leq Q^2 \equiv (\Ptg)^2 \leq 1.6 \times 10^{5} ~\rm GeV^2 $, 
extending the kinematic reach of previous photon plus jet 
measurements (see refs. inside \cite{gamjet_PLB}). Ratios between the differential 
cross sections in the four studied angular configurations are also presented. The measurements are 
compared to the corresponding theoretical predictions.

The data presented here correspond to an integrated luminosity of 
1.01~$\pm$~0.06 fb$^{-1}$ collected using the D0 detector~\cite{D0_det}
at the Fermilab Tevatron $p\bar{p}$ collider operating at a center-of-mass energy $\sqrt{s}=$1.96 TeV. 

Events containing at least one hadronic jet are selected. Jets are reconstructed using the 
D0 Run II jet-finding algorithm with a cone of radius 0.7 \cite{Run2Cone}, 
and are required to satisfy 
quality criteria which suppress background from leptons, photons, and 
detector noise effects. 
Jet energies are corrected to the particle level.

In total, about 1.4 million candidate events are selected after application 
of all selection criteria. A correction for the ``$\gamma$+jet'' event purity ${\cal P}$
is then applied to account for the remaining background in the region $O_{\rm NN}>0.7$.
The distribution of the ANN output for the simulated 
photon signal and dijet background samples are fitted to 
the data for each $\Ptg$ bin using a maximum likelihood fit
\cite{HMCMLL} to obtain the fractions of signal and background components.
The found purities vary as 0.52 -- 0.99 for $30 < \Ptg < 400$ GeV.
%

The data are compared to next-to-leading order (NLO) QCD
predictions obtained using {\sc jetphox} \cite{JETPHOX}, with 
CTEQ6.5M PDF \cite{CTEQ} and BFG fragmentation functions of partons to photons \cite{BFG}. 
The renormalization, factorization, and 
fragmentation scales ($\mu_{R}$, $\mu_{F}$, and $\mu_f$) are set equal to
$\newscale$, where $f({y^\star})=\Fystar$ and $y^\star=\ystar$ \cite{Scale_choice}.

The prediction using the CTEQ6.5M PDF and BGF fragmentation sets does not 
describe the shape of the cross section over the whole measured range. 
In particular, the prediction is above the data for events with $|y^{\rm jet}|<0.8$ 
in the region $\Ptg \gt 100$ GeV and below the data for jets produced in the 
$1.5\lt |y^{\mathrm{jet}}| \lt 2.5$, $y^{\mathrm{\gamma}} \cdot y^{\mathrm{jet}}>0$ 
rapidity region for $\Ptg \lt 50$ GeV. 
Most of the data points in these $\Ptg$ and rapidity regions are 
(1--1.5)~$\delta \sigma_{\rm tot}$ outside of the CTEQ6.5M PDF set uncertainty range which is 
shown by the shaded region in the figure and calculated according to 
the prescription in \cite{CTEQ}. 
Note that the  data-to-theory ratios have a shape similar to those
observed in the inclusive photon cross sections measured by the UA2~\cite{UA2}, 
CDF \cite{CDF} and D0 \cite{IncPhotons} collaborations.

%
\begin{figure}[t]
\includegraphics[scale=0.42]{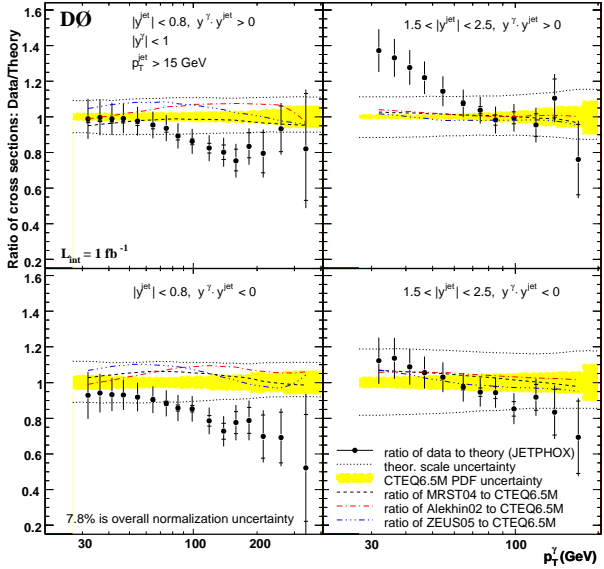} 
\caption{The ratios of the measured triple-differential cross section, in each measured interval, 
to the NLO QCD prediction using {\sc jetphox} \cite{JETPHOX}
with the CTEQ6.5M PDF set and all three scales $\mu_{R,F,f}=p_T^\gamma f({y^\star})$.
The solid vertical line on the points shows the statistical and $p_T$-dependent systematic uncertainties added in quadrature, 
while the internal line shows the statistical 
uncertainty. The two dotted lines represent the effect of varying the 
theoretical scales by a factor of two. 
}
\label{fig:DT_reg1}
~\\[-6mm]
\end{figure}
\begin{figure}[t]
\includegraphics[scale=0.42]{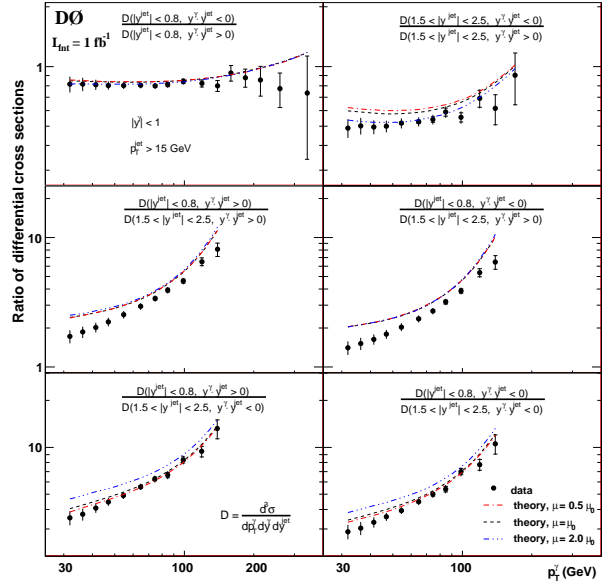}
\caption{The ratios between the differential cross sections in 
each $y^{\mathrm{jet}}$ region. The solid vertical error bars correspond to the statistical
and systematic uncertainties added in quadrature while the
horizontal marks indicate the statistical uncertainty. NLO QCD theoretical predictions 
for the ratios, estimated using {\sc jetphox}, are shown for three different 
scales: $\mu_{R,F,f}$=$\mu_{0}$, 0.5$\mu_{0}$, and 2$\mu_{0}$, where $\mu_{0}=p_T^\gamma f({y^\star})$.}
\label{fig:cross_ratio1}
~\\[-4mm]
\end{figure}

The experimental systematic uncertainties are reduced further by measuring
the ratios between the differential cross sections 
$\mathrm{D}=\mathrm{d^3}\sigma / \mathrm{d}\Ptg\mathrm{d}y^{\gamma}\mathrm{d}y^{\mathrm{jet}}$ 
in the different regions. Most of the systematic uncertainties related to the 
identification of central photons then cancel, and only systematic 
uncertainties related to the \gpj ~event purities and the jet selection 
efficiency (when measuring ratios between central and forward jet 
regions) remain. 
Measured ratios between the differential cross sections 
in the different regions are presented in Fig.~\ref{fig:cross_ratio1}.
%
%
The ratios also significantly reduces theoretical scale uncertainties \cite{gamjet_PLB}.
The shapes of the measured ratios between the cross sections in the 
different regions, in general, are qualitatively reproduced by the
theory. A quantitative 
difference, however, between theory and the measurement is observed for the 
ratios of the central jet regions to the forward $1.5\lt |y^{\mathrm{jet}}|\lt2.5$, $y^{\mathrm{\gamma}}\!\cdot\! y^{\mathrm{jet}}>0$ 
region, even after the theoretical scale variation is taken into account.  The ratio between the 
two forward jet cross sections suggests a scale choice $\mu_{R,F,f} \simeq 2 p_T^\gamma f({y^\star})$. 
However, the ratios of the central jet regions to the forward 
$1.5\lt |y^{\mathrm{jet}}|\lt2.5$, 
 $y^{\mathrm{\gamma}}\!\cdot\! y^{\mathrm{jet}}<0$ region suggest a 
theoretical scale closer to $\mu_{R,F,f} \simeq 0.5 p_T^\gamma f({y^\star})$.
%


\section{Measurement of $\gamma+b+X$ and $\gamma+c+X$ production 
cross sections in $p\bar{p}$ collisions at $\sqrt{s}=~1.96$~TeV}

\label{sec:gamHFjet}

Photons produced in association with heavy quarks $Q$
($\equiv c$ or $b$) in the final state of hadron-hadron interactions
provide valuable information about the parton distributions of the
initial state hadrons~\cite{Berger_charm,CTEQ_c}.  
Such events are produced primarily through the QCD Compton-like scattering process
$gQ\to \gamma Q$ 
but also through quark-antiquark annihilation $q\bar{q}\to
\gamma g \to \gamma Q\bar{Q}$.  Consequently, $\gamma + Q+X$
production is sensitive to the $b$, $c$, and gluon ($g$) densities
within the colliding hadrons, and can provide constraints on parton
distribution functions  that have substantial
uncertainties~\cite{CTEQ}.  The heavy quark and gluon content
is an important aspect of QCD dynamics and of the fundamental
structure of the proton.  In particular, many searches for new
physics, e.g. for certain Higgs boson production
modes~\cite{Diff_Higgs,Ch_Higgs,Higgs_rev}, will benefit
from the increased PDF precision with respect to the heavy quark and
gluon content of the proton.

This section presents the first measurements of the inclusive
differential cross sections \triplexs for $\gamma+b+X$ and
$\gamma+c+X$ production in $p\bar{p}$ collisions, where $y^\gamma$ and
$y^{\rm jet}$ are the photon and jet rapidities \cite{gamjet_PRL}.  The
results are based on an integrated luminosity of
1.02$~\pm~0.06$~fb${}^{-1}$  collected with the D0
detector at the Fermilab Tevatron Collider at
$\sqrt{s}=$ 1.96 TeV.  The highest $p_T$ (leading) photon and jet are
required to have $|y^{\gamma}|<1.0$ and $|y^{\rm jet}|<0.8$, and
transverse momentum $30<\Ptg< 150$~GeV and $p_{T}^{\rm jet}>15$~GeV.
This selection allows one to probe PDFs in the range of
parton-momentum fractions ($x$) $0.01\lesssim x\lesssim 0.3$, and hard
scatter scales of $9\times 10^2 \lesssim Q^2 \equiv
(p_{T}^{\gamma})^2\lesssim 2\times 10^4~\GeV^2$.  Differential cross
sections are presented for two regions of kinematics, defined by
$y^{\gamma} \cdot y^{\rm jet}>0$ and $y^{\gamma} \cdot y^{\rm jet}<0$.  These
two regions provide greater sensitivity to the parton $x$ because they
probe different sets of $x_1$ and $x_2$ intervals, as discussed in
\cite{gamjet_PLB}.

Photon selection criteria and methods for estimating photon purities
are similar to those used in \cite{gamjet_PLB}.
At least one jet must be present in each event. Jets are reconstructed
using the D0 Run~II algorithm~\cite{Run2Cone} with a radius of $0.5$.  
Light jets are suppressed
using a dedicated artificial neural network ($b$-ANN)~\cite{bNN}
that exploits the longer lifetimes of heavy-flavor hadrons relative to
their lighter counterparts. The leading jet is required to have a
$b$-ANN output value $>0.85$.  

About 13,000 events remain in the data sample after applying all
selection criteria.  
To estimate the
photon purity, the $\gamma$-ANN distribution in data is fitted to a
linear combination of templates for photons and jets obtained from
simulated $\gamma~+$ jet and dijet samples, respectively.  An
independent fit is performed in each $\Ptg$ bin, yielding photon
purities between 51\% and 93\% for $30<\ptg<150~\GeV$.  The fractional
contributions of $b$ and $c$ jets are determined by fitting templates
of $\rJLIP=-\ln\prod_{i}{P_{\rm track}^{i}}$ to the data, where
$P_{\rm track}^{i}$ is the probability that a track originates from
the primary vertex, based on the significance of the track's distance
of closest approach to the primary vertex.  
Templates are used for the shape information
of the $\rJLIP$ distributions.  For $b$ and $c$ jets these are
extracted from MC events while the light jet template is taken from a
data sample enriched in light jets, which is corrected for
contributions from $b$ and $c$ quarks.
\begin{figure}[t]
\includegraphics[width=5.5cm,height=5cm,trim=27 30 55 50,clip=true]{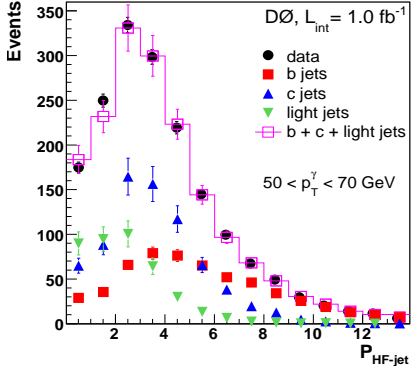}
\caption{Distribution of observed events for $\rJLIP$
  after all selection criteria for the bin $50<\Ptg< 70$~GeV.  The
  distributions for the $b$, $c$, and light jet templates are shown
  normalized to their fitted fraction.  Error bars on the templates
  represent combined uncertainties from statistics of the MC and the
  fitted jet flavor fractions, while the data contain just statistical
  uncertainties.  Fits in the other $\Ptg$ bins are of similar quality.}
\label{fig:cbjet_test}
\end{figure}
\begin{figure}[t]
\includegraphics[width=5.5cm,height=5cm,trim=10 20 55 50,clip=true]{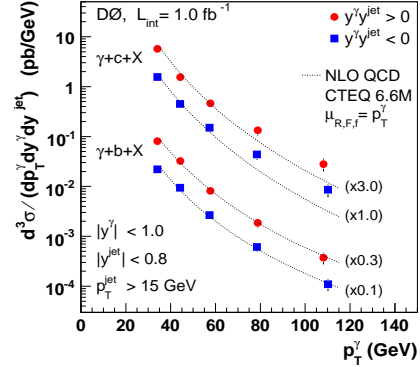}
\caption{ The \gb and \gc differential cross sections
  as a function of $\Ptg$ in the two regions $y^{\gamma}
  y^{\rm jet}>0$ and $y^{\gamma} y^{\rm jet}<0$.  The uncertainties
  on the data points include statistical and systematic contributions
  added in quadrature. The NLO pQCD predictions using {\sc cteq}6.6M
  PDFs are indicated by the dotted lines.}
\label{fig:xsectbc1plot}
\end{figure}

The result of a maximum likelihood fit, normalized to the number of
events in data, is shown in Fig.~\ref{fig:cbjet_test} for
$50<\Ptg<70~\GeV$, as an example.  The estimated fractions of $b$ and $c$ jets in all
$\Ptg$ bins vary between 25--34\% and 40--48\%, respectively.  The
corresponding uncertainties range between
7--24\%, dominated at higher $\Ptg$ by the limited data statistics.

The measured differential cross sections are shown in
Fig.~\ref{fig:xsectbc1plot} for \gb and \gc production as a function
of $\Ptg$ for the jet and photon rapidity intervals in question.  The
cross sections fall by more than three orders of magnitude in the
range $30<\Ptg< 150~\GeV$.  
The statistical uncertainty on the results
ranges from 2\% in the first $\Ptg$ bin to $\approx 9\%$ in the last
bin, while the total systematic uncertainty varies between 15\% and
28\%.  

\begin{figure}
\includegraphics[width=\linewidth]{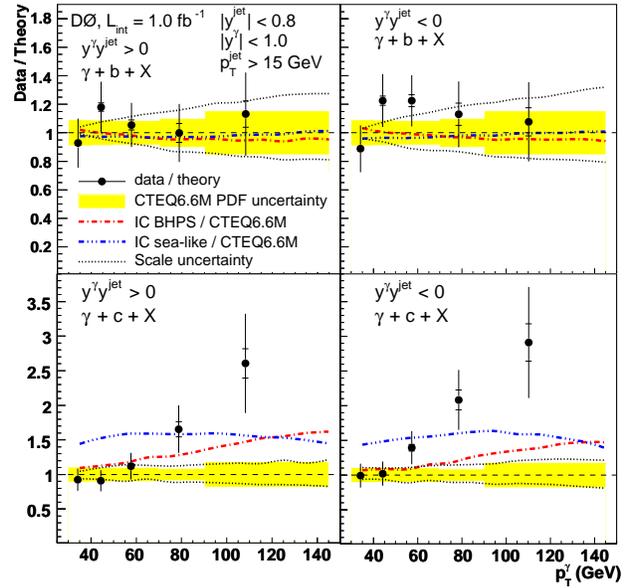}
\caption{The data-to-theory ratio of cross sections as a function of
  $\Ptg$ for \gb and \gc in the regions $y^{\gamma} y^{\rm jet}>0$
  and $y^{\gamma} y^{\rm jet}<0$.  The uncertainties on the data
  include both statistical (inner line) and full uncertainties (entire
  error bar).  Also shown are the uncertainties on the theoretical
  pQCD scales and the {\sc cteq}6.6M PDFs.  The scale uncertainties
  are shown as dotted lines and the PDF uncertainties by the shaded
  regions.  The ratio of the standard {\sc cteq}6.6M prediction to two
  models of intrinsic charm is also shown.}
\label{fig:xsectb1ratio}
\end{figure}

NLO pQCD predictions \cite{Tzvet}, with
the renormalization scale $\mu_{R}$, factorization scale $\mu_{F}$,
and fragmentation scale $\mu_f$, all set to $\Ptg$,
are compared to data in Fig.~\ref{fig:xsectbc1plot}.  
The ratios of the measured to the
predicted cross sections are shown in Fig.~\ref{fig:xsectb1ratio}.

The uncertainty from the choice of the scale is estimated through a
simultaneous variation of all three scales by a factor of two, i.e.,
to $\mu_{R,F,f}=0.5 p_T^\gamma$ and $2 p_T^\gamma$.  The predictions
utilize {\sc cteq}6.6M PDFs~\cite{CTEQ}, and are corrected for effects
of parton-to-hadron fragmentation.  This correction for $b\,(c)$ jets
varies from $7.5$\% ($3$\%) at $30<\Ptg<40~\GeV$ to 1\% at
$90<\Ptg<150~\GeV$.

The pQCD prediction agrees with the measured cross sections for \gb
production over the entire $\ptg${} range, and with \gc production for
$\ptg<70$~GeV.  For $\ptg>70$~GeV, the measured \gc cross section is
higher than the prediction by about 1.6--2.2 standard deviations
(including only the experimental uncertainties) with the difference
increasing with growing $\ptg$.

Parameterizations for two models containing intrinsic-charm (IC) have
been included in {\sc cteq}6.6 \cite{CTEQ_c}, and their ratios to the
standard {\sc cteq} predictions are also shown in
Fig.~\ref{fig:xsectb1ratio}.  Both non-perturbative
models predict an increased \gc cross section, and for the BHPS
model~\cite{CTEQ_c} it grows with $\Ptg$.  The observed difference may
also be caused by an underestimated contribution from the $g\to
Q\bar{Q}$ splitting in the annihilation process that dominates for
$\Ptg>90~\GeV$~\cite{PDG}.

\section{Double parton interactions in $\gamma$ + 3 jet events in $p\bar{p}$ collisions 
at $\sqrt{s}=1.96$~TeV in \DO} 

\label{sec:DP}

   Many features of high energy inelastic hadron collisions 
  depend directly on the parton structure of hadrons which is 
  still not yet well understood at both the theoretical and
  experimental levels.
  Phenomenologically, the proton (or antiproton)
  may be  viewed, in the first naive  approximation, 
  as an object composed of three light quarks (or anti-quarks). The study of this 
  structure is founded mainly on
  the use of a simplified theoretical model  which 
  considers high energy inelastic scattering of nucleons    
  as a process involving a single collision of one quark or gluon from one nucleon with one
  quark or gluon from the other nucleon. In this approach, additional
  ``spectator'' partons do not take part in the hard
  $2 \to 2$ parton collision and form the so-called 
  ``underlying event''. 

  Another, much less developed, approach is based on models 
  in which there might be more than one hard interaction of parton pairs in one collision of nucleons. 
  Since each incoming hadron is a composite object, consisting of many partons,
  such a probability should be non-zero.
  Models with  multiple parton collisions have been
  considered in a few theoretical papers (see refs. inside \cite{DP_CONF}). 
    It is obvious that the rate of events with double parton scattering (DPS)
   depends on how the partons are distributed within 
   the nucleon. The form of the parton spatial distribution 
   and possible correlations between partons
   is practically unknown. This information is hard to obtain
   within the present theoretical models based on perturbative
   QCD and makes a relevant measurement  particularly important.

  Recently the  information about a fraction of the double parton 
  interactions has become very important due to growing Tevatron
  luminosity and upcoming LHC experiments.
  It opens the possibility to 
  search for signals from new physics processes for which the DPS events may give a noticeable background. 

  So far, there have been only four dedicated  measurements studying double parton scattering:
   the AFS experiment in $pp$ collisions at $\sqrt{s}=63$ GeV \cite{AFS_dp},
   UA2 in $p\bar{p}$ collisions at $\sqrt{s}=630$ GeV \cite{UA2_dp}
   and two by CDF in $p\bar{p}$ collisions $\sqrt{s}=1.8$ TeV \cite{CDF93,CDF97}.

   In the present analysis, we use a sample of \gpTHRj events collected
   by the D0 experiment during Run~IIa at the integrated luminosity of 1 fb$^{-1}$
   in $p\bar{p}$ collisions at $\sqrt{s}=1.96$ TeV \cite{DP_CONF}.
   We determine the fraction of the double parton interactions in a single $p\bar{p}$ collision 
   and also the value of ``effective cross section'' $\sigma_{\rm eff}$. The latter allows,
   with the known cross sections $\sigma^{A}$ and $\sigma^{B}$ for two independent parton scatterings
   A and B, to calculate $\sigma_{\rm DPS}$ cross section: \\[-5mm]
   \begin{eqnarray}
   \sigma_{DPS} \equiv
     \frac{\sigma^{A} \sigma^{B}}{\sigma_{\rm eff} }~
   \end{eqnarray} 
   Here  the normalization factor $\sigma_{\rm eff}$ is a parameter
   that can be directly related to the distance
   between partons in the nucleon. 
   If the partons are uniformly distributed inside the 
   nucleon (large $\sigma_{\rm eff}$), $\sigma_{DPS}$ will be rather low and, conversely, 
   it should increase for a highly concentrated parton spatial densities (small $\sigma_{\rm eff}$).
   A better energy measurement of photons as compared with jets
   helps in separating the DP scatterings and allows us to better fix the scale of the main
   hard interaction.

%
%
    In the current analysis, to extract $\sigma_{\rm eff}$ we use a technique
    that operates only with quantities determined from  data analysis and minimizes many theoretical assumptions 
    \cite{CDF97,DP_CONF} that were used in some previous measurements. 
%
    Towards this aim, we measure the number of DP \gpTHRj events and the number of \gpTHRj events with hard interactions
    occurring in two separate $p\bar{p}$ collisions. The latter class of events will be called
    double interactions (DI).  
    Assuming independent (uncorrelated) scatterings in the DP process \cite{Sjost,TH2}, the DP and DI
    events  should be kinematically identical and differ by just amount of ``underlying'' energy
    in one- and two-vertex events.
%
%
%
   The expression for the effective cross section $\sigma_{\rm eff}$  can be written  in the following form: \\[-3mm]
   \begin{eqnarray}   
   \sigma_{\rm eff} = 
  \frac{N_{\rm DI}}{N_{\rm DP}} \frac{N_{c}(1) } {2N_{c}(2)} 
   \sigma_{\rm hard}. 
   \label{eq:sig_eff}
   \end{eqnarray}
  Here  $N_{\rm DI(DP)}$ is the number of DI(DP) events, $ N_{c}(1)$ and  $ N_{c}(2)$ are the numbers
  of beam crossings with 1 and 2 hard collisions,
   and $\sigma_{\rm hard}$ is the cross section of the hard $p\bar{p}$ interaction.
   We measure $\sigma_{\rm eff}$ in three bins of the 2nd (ordered in $p_T$) jet ($\ptsj$) 
   which serves as a scale of the additional parton-parton interaction.


%
We used data collected with the D0 detector during Run~IIa
which 
corresponds an integrated luminosity of about 1.02 $\pm$ 0.06 fb$^{-1}$.
The data sets and $\gamma$+jets selection criteria, analogous to the previous \gpj measurements 
\cite{gamjet_PLB,gamjet_PRL},
have been used in this analysis. 


   The main background for the DP events is caused by the $\gamma + 3$ jets events
   resulting from the single parton (SP) scatterings with hard gluon radiation
   in the initial or final state $qg \to q\gamma gg$ or $q\bar{q} \to g\gamma gg$.
   The fraction of the DP events is determined in this analysis using a set of variables 
   sensitive to the kinematic
   configurations of the two independent scatterings of parton pairs.
specifically to the difference between between the $p_T$ imbalances of two pairs in \gpTHRj events:\\[-1mm]
\begin{eqnarray}
S_{p_{T}}=\frac{1}{\sqrt{2}}
\sqrt{{\left(\frac{\left|\vec{p}_{T}(\gamma,i)\right|}{\delta p_{T}(\gamma,i)}\right)}^{2}+
{\left(\frac{\left|\vec{p}_{T}(j,k)\right|}{\delta p_{T}(j,k)}\right)}^{2}}
\label{eq:S_var}
\end{eqnarray}

%
Two other variables, similar to (\ref{eq:S_var}), have also been used in the analysis \cite{DP_CONF} for a cross-check. 
In the equations above, $\vec{p}_{T}(\gamma,i)$ and $\vec{p}_{T}(j,k)$ are the $p_T$ vectors of 
the total transverse momenta of the two two-body system, 
and $\delta p_{T}(\gamma,i), \delta p_{T}(j,k)$
are the corresponding uncertainties. 
The pairs are constructed by grouping $\gamma$ with 3 jets in three possible configurations.
%
The configuration that gives the minimum $S$ is selected for each of the $S$-family variables.
In most events 
$S$ is minimized by pairing the photon with the leading jet, while the additional jets both come
from dijet system or one of them is replaced by the radiation jet.

The $\Delta S$-family variables 
are allied to the $S$-family. They are defined
as an azimuthal angle between the $p_{T}$ vectors of the pairs that give a minimum $S$ value:\\[-3mm]
\begin{eqnarray}
\Delta S =\Delta\phi\left(\vec{p}_T^{\gamma, jet_i}, ~\vec{p}_T^{jet_j, jet_k}\right)
\label{eq:DeltaS_var}
\end{eqnarray}
where $\vec{p}_T^{\gamma, jet_i}$ and $\vec{p}_T^{jet_j, jet_k}$ are 
the total $p_T$ balance vectors of the pairs ($\gamma+jet_i$)
and  ($jet_j+jet_k$) which give minimal $S$.

The models for signal DP (and DI) events have been constructed directly from data mixing
$\gamma+\geq$1 jet and dijet minimum bias events in the one- (and two-) vertex data \cite{DP_CONF}.
Then the fractions of DP events have been found from analysis of shapes for $\Delta S$ 
distributions in the signal events and data in the adjacent $\ptsj$ bins 
for which we should expect different DP fractions \cite{TH2}.
Fractions of DI events have been found fitting the shapes for $\Delta S$ distributions
of signal and background events to data.
The obtained DP fractions are shown in Tables \ref{tab:dp_frac}, 
while the DI fractions vary as   $0.189 \pm 0.029$ at $15<\ptsj<20$ GeV,
$0.137\pm 0.027$ at $20<\ptsj<25$ GeV and $0.094 \pm 0.025$ at $25<\ptsj<30$ GeV.

\begin{table}[htpb]
\vskip -0mm
\begin{center}
\small
\caption{Fractions of DP events $f_{\rm DP}$ found for the three $\ptsj$ intervals (GeV).}
\label{tab:dp_frac}
\vskip 1mm
\begin{tabular}{|c|c|c|c|} \hline
~$\ptsj$~ & $15-20$ & $20-25$ & $25-30$ \\\hline
~$f_{\rm DP}$~ & 0.466$\pm$0.041 ~&~   0.334$\pm$0.023 ~&~   0.235$\pm$0.027 \\\hline
\end{tabular}
\end{center}
\end{table}


%

\begin{table}[htpb]
\vskip -7mm
\begin{center}
\small
\caption{Effective cross section $\sigma_{\rm eff}$ (mb) found in the three $\ptsj$ intervals (GeV).}
\label{tab:sigma_eff}
\vskip 1mm
\begin{tabular}{|c|c|c|c|} \hline
~$\sigma_{\rm eff}$~ &  $15-20$ & $20-25$ & $25-30$ \\\hline
 $\ptsj$ & $16.2 \pm 2.8$ ~&~ $13.8 \pm 3.1$  ~&~ $11.5 \pm 4.7$  \\\hline
\end{tabular}
\end{center}
\vskip -2mm
\end{table}

The resulting values of $\sigma_{\rm eff}$ with total
(systematic $\oplus$ statistics) uncertainties are given in Table \ref{tab:sigma_eff} for the three $\ptsj$ bins.
%
The total systematic uncertainty varies between 19\% and 31\% and
is mainly caused by the determination of the DI and DP fractions as well as by the ratios of the DP/DI selection efficiencies.

One can see that the obtained $\sigma_{\rm eff}$ values in different jet $p_T$ bins 
agree with each other within their uncertainties.
Using this fact and also that the uncertainties in different jet $p_T$ bins have very small correlation,
we can calculate the $\sigma_{\rm eff}$ value averaged over the three jet $p_T$ bins.
It gives us \\[-4mm]
\begin{equation}
 \sigma_{\rm eff}^{aver.} = 15.1 \pm 1.9 {~~\rm mb}. 
  \label{eq:sigeff_av}
\end{equation}

%
      It is worth mentioning that the obtained average value is
      in the range of those found in previous analogous measurements \cite{UA2_dp}--\cite{CDF97}.
      This closeness  may indicate a tendency
      for a stable behavior of $\sigma_{\rm eff}$ with respect to the
      transverse momentum of the jet produced in the second parton-parton interaction.

\section{Conclusion}
The presented measurements of production cross sections for 
$\gamma+$jet, $\gamma+$heavy flavour jet and study of the events with double parton scatterings
provide precision tests of perturbative QCD as well as information on the fundamental
structure of the nucleon; momentum and spatial distributions of partons in the nucleon.
In addition to these important aspects of QCD dynamics, this knowledge 
should be very helpful for many searches for new physics since, for example,
$\gamma+$jet and $\gamma+b(c)$ jets are noticeable components for the background to
many Higgs boson decay modes (e.g. $h\to \gamma\gamma, bb$) as well to extra dimensions
with graviton ($G\to ee,\gamma\gamma$) and technicolor (e.g. $\omega_T\to \gamma\pi_0$, $a_T^\pm\to \gamma\pi^\pm$
with $\pi^0 \to b\bar{b}$, $\pi^\pm \to b\bar{c}, \bar{b}c$) \cite{TC}.
Knowledge of parton matter structure and parton correlations being extremely valuable {\it per se}
is also important for understanding multijet production characteristic for many physics final states,
especially expected in many SUSY models \cite{PDG}.





\end{document}